\documentclass{article}
\def \inbar{\vrule height1.5ex width.4pt depth0pt}
\def \C{\relax\hbox{\kern.25em$\inbar\kern-.3em{\rm C}$}}
\def \R{\relax{\rm I\kern-.18em R}}

\newcommand{\sgn}{{\rm sgn}}

\newcommand{\beq}{\begin{equation}}
\newcommand{\eeq}{\end{equation}}
\newcommand{\bea}{\begin{eqnarray}}
\newcommand{\eea}{\end{eqnarray}}
\newcommand{\nn}{\nonumber}

\newcommand{\pdr}{\partial}

\newcommand{\Tr}{\hbox{Tr}}

\begin{document}
\author{O. Teoman Turgut${\,}^{1,2,3}$\\
 \\
${}^{1}\,$Department of Physics,  Bogazici University  \\
80815, Bebek, Istanbul, Turkey\\
\\
${}^{2}\,$Department of Physics, KTH\\
SE-106 91 Stockholm, Sweden\\ 
\\
${}^{3}\,$Feza Gursey Institute\\
 81220 Kandilli, Istanbul, Turkey\\
  turgutte@boun.edu.tr\\
 turgut@theophys.kth.se}

\title{\bf On two  dimensional coupled bosons and fermions }
\maketitle
\large
\begin{abstract}
\large
We study complex bosons and fermions coupled through a 
generalized Yukawa type coupling in the large-$N_c$ limit
 following ideas of Rajeev[Int. Jour. Mod. Phys. {\bf A 9} (1994) 5583].  
We study  a linear approximation to this model. We show that in this 
approximation we do not have boson-antiboson 
and fermion-antifermion bound states occuring together. 
There is a possibility of having only fermion-antifermion bound 
states. We support this claim by finding distributional 
solutions with energies lower than the two mass treshold in the fermion
sector. This has important implications from 
the point of view of scattering  theory.
We discuss some aspects of the scattering above the two mass 
treshold of   boson pairs  and  fermion pairs.
We also briefly present  a gauged version
of the same  model and  write down the linearized 
equations of motion.
\end{abstract}
\large
\section{Introduction}

Quantum field theories are both fundamental and challenging.
Despite the fact that our description of the world of elementary
particles is based on  quantum field theory,
we still do not have  a complete  understanding of interacting 
field theories, especially their bound state structure. 
For this reason it is interesting to 
study simple examples where one can actually 
make more  progress.
There are various such models in two dimensions and 
they have been a valuable source for new  ideas and 
testing ground for many  years (see 
\cite{abdalla} for a comprehensive selection of topics).

In this article we study a two dimensional model which could be
another possible toy model for understanding Yukawa coupled 
field theories in four dimensions.
The physically important one is  the  gauged  Yukawa theory, as 
we know from
the present day version of the standart model.
In this work we  will also study some aspects of 
the gauge coupled interacting bosons and fermions.
We will actually present the linearized equations 
of the gauged version at the end of our
paper, but our main emphasis is to understand the
simpler case without the gauge potentials (of course 
it is not so clear if this is really a simpler theory).
A more complicated version of the model we discuss 
is  investigated in \cite{cavicchi} using path integral techniques.
The  results of the present  paper are somewhat  different
since we follow a Hamiltonian approach (and 
we are not taking the most complicated possible version).

Let us comment on  some fundamental work in 
the literature that we are aware of:
the literature on Yukawa theory is vast, we will mention only
a few of them, a rigorous construction  of two dimensional model is 
given in the papers \cite{glimm, schrader}, the contstruction of the 
proability measure within the Euclidean formalism 
is done  in \cite{seiler}. The most recent rigorous analysis was given in 
\cite{lesniewski} by following a renormalization group type idea
essentially inspired from \cite{gawedzki}.
It will be interesting to attempt such a  rigorous approach 
for the model we discuss below.
The standart 
Yukawa coupling in two dimensions in the light-cone approach is discussed in 
\cite{pauli1, pauli2}. A further analysis using 
the Tamm-Dancoff approximation in the light-cone 
is pursued in \cite{perry}. 
A very interesting discussion of the Yukawa model in 
four dimensions is presented in \cite{glazek}, clearly 
a four dimensional model has many more interesting features.
A more recent analysis of the fermion  bound states 
of the same model is discussed in \cite{brinet}.
The equivalence between 
the light-cone and covariant perturbation theory 
is analysed in \cite{baker}.
(Good reviews of field theories in the light cone 
are given in \cite{brodsky} and 
renormalization of light cone Hamiltonians in \cite{perry2}).

In this work, we introduce a color degree of freedom 
for the purpose of reaching a Hartree-Fock type 
approximation. Following the ideas suggested by 
Rajeev in \cite{2dqhd} we reformulate the 
 problem in terms of color invariant bilinears
(see also \cite{wadia} for some similar ideas).
The details of this refromulation  are explained in our 
previous work \cite{tolyateo} within the context 
of gauge theories, therefore 
in this work we will use the results of the cited 
article directly. In some sense the present article 
is a natural continuation.
The reader should consult \cite{tolyateo} for 
more information on the geometry of the 
resulting classical phase space.
We also recommend the lectures notes of Rajeev on 
two dimensional QCD \cite{istlect}.
In \cite{2dqhd}, it is shown that the 1+1 dimensional 
QCD in the large-$N_c$ limit can be reformulated
as a classical field theory with an infinite dimensional
phase space, which is identified to be the 
restricted Grassmanian.  
The study of two dimensional QCD in the large-$N_c$ limit 
is given by 't Hooft in his well-known paper \cite{thooft},
its generalization to scalar fields is done in 
\cite{shei} and in \cite{tomaras} using Hamiltonian 
methods. The two dimensional combined fermions and bosons  QCD  
is given by Aoki in \cite{aoki1, aoki2} and also discussed by 
Cavicchi \cite{cavicchi}.
Within Rajeev's approach one can reach the same results 
by using a linear approximation to the full theory.
One can further study baryons by using a variational 
ansatz which does not correspond to small fluctuations 
of the fields, therefore cannot be seen by the 
linear approximation. 
Here we obtain the general Hamitonian in the 
large-$N_c$ limit for gauge coupled bosons and fermions 
which are also interacting through a generalized Yukawa type
interaction. The meson equations are  given for the linearized 
theory. Our presentation is incomplete since 
we do not study beyond the linear approximation and we plan to 
come to a more detailed analysis in the future.

We note that in the simpler model we discuss one can actually 
solve the integral equations, ending up with some 
eigenvalue or scattering solutions.
These equations require a simple renormalization to
be meaningful (which at the end amounts to defining the 
singular integrals as the Hadamard principal value).
We warn the reader that the form of the resulting Hamiltonian suggests that 
a physically more relevant approximation in the
non-gauged models could be given by 
a variational ansatz. This is due to the fact that the 
interactions in the 
linear approximation are all multiplied by the 
fermion mass. For heavy fermions we expect that the
linear approximation gives valuable results, but 
for example in the case of massless fermions
all the information is contained in the 
higher order terms, which cannot be accessed by 
the method we use.

\section{Coupling between  Complex Bosons and Fermions}

We start with the action of our model with two Yukawa type couplings,
\beq
S=\int d^2x\Big( i\bar \psi^\alpha \gamma^\mu \partial_\mu \psi_\alpha
-\bar \psi^\alpha(\mu_{1Y}\phi^{*\beta}\phi_\alpha+
\mu_{2Y}\phi^{*\lambda}\phi_\lambda\delta^\alpha_\beta+
 m_F\delta_\alpha^\beta)\psi_\beta
+\partial^\mu\phi^{*\alpha}\partial_\mu\phi_\alpha-
m_{B0}^2\phi^{*\alpha}\phi_\alpha-{\lambda^2_{B0}\over 4}(\phi^{*\alpha}
\phi_\alpha)^2\Big)
.\eeq
Here $\alpha$ refers to a common flavor index and it 
goes from $1$ to $N_c$ (we continue to write it as color symmetry, since 
at the end we will also talk about the gauged model).
It is more natural to keep $\mu_{1Y}=\mu_{2Y}$ when there are no gauge fields,
since we will use the color degrees as a way of reaching a mean-field 
description, but we will keep this more general form.
We rewrite the action  in the light-cone coordinate system,
and {\it  we choose $x^+$ as our ``time'' coordinate}, 
\bea
 S&=&\int dx^+dx^-\Big(i\sqrt{2}\psi^{*\alpha}_L\partial_-\psi_{L\alpha}
+i\sqrt{2}\psi^{*\alpha}_R\partial_+\psi_{R\alpha}+
 \phi^{*\alpha}(-2\partial_-)\partial_+\phi_\alpha-
m_{0B}^2\phi^{*\alpha}\phi_\alpha\nn\cr
&-&(\psi^{*\alpha}_L\psi_{R\beta}+\psi^{*\alpha}_R\psi_{L\beta})
(\mu_{1Y}\phi^{*\beta}\phi_\alpha+\mu_{2Y}\phi^{*\lambda}\phi_\lambda
\delta^\alpha_\beta+m_F\delta_\alpha^\beta)
-{\lambda_{B0}^2\over 4}(\phi^{*\alpha}
\phi_\alpha)^2\Big)
.\eea
There are many good introductions to the light-cone field theory,
we refer the reader to \cite{heinzl, yan, harindra}.
We note that the lefthanded components of the fermion
field  are  nondynamical, therefore we will 
 remove
$\psi_{L\alpha}$ and its complex conjugate through the equations of 
motion in the quantum theory
(we refer to \cite{tolyateo} for our conventions in quantizing
this theory, below we summarize the results),
\beq
 \hat \psi_{L\alpha}={\sqrt{2}\over 2i\partial_-}[\mu_{1Y}:\hat \phi_\alpha
\hat \phi^{\dag\beta}\hat \psi_{R\beta}:+\mu_{2Y}:\hat \phi^{\dag\beta}\hat
\phi_\beta\hat \psi_{R\alpha}:+
m_F\hat\psi_{R\alpha}]
.\eeq
We do not really need to worry about the normal ordering in the first
factor, since
in the large-$N_c$ limit these corrections will be of smaller 
order, we wrote it to emphasize that the reduction should be performed 
at the quantum level. At the second term we have a normal ordering for the 
color contracted bosons only. 
The second thing we notice from 
the light-cone action is that we are already in 
the Hamiltonian formalism. Therefore 
we can read off the Hamiltonian directly from 
the action when  we insert the solution of the nondynamical field 
 back into the action. Thus   we arrive at the following 
Hamiltonian,
({\it from now on we write $\psi$ for $\psi_R$ since 
this is the only fermionic field we have}), 
\bea
  \hat H&=&\int dx^-\Big( {\sqrt{2}m_F^2\over 2}:\hat \psi^{\dag\alpha}
{1\over i\partial_-}\hat \psi_\alpha:+ m_{B0}^2:\hat \phi^{\dag\alpha}
\hat \phi_{\alpha}:+{\lambda^2_{B0}\over 4}
(:\hat\phi^{\dag\alpha}\hat\phi_\alpha:)^2\nn\\
&\ & +{\sqrt{2}\over 2} \mu_{1Y}m_F[\hat \psi^{\dag\alpha}{1\over i\partial_-}
\hat \phi_\alpha\hat \phi^{\dag \beta} \hat \psi_\beta+
\hat \psi^{\dag\beta}\hat \phi_\beta\hat\phi^{\dag\alpha}{1\over i\partial_-}
\hat\psi_\alpha]+
{\sqrt{2}\over 2}\mu_{1Y}^2\hat\psi^{\dag\beta}\hat\phi_\beta
\hat\phi^{\dag\alpha}
{1\over i\partial_-}\hat\phi_\alpha\hat \phi^{\dag\lambda} 
\hat \psi_\lambda\nn\\
&\ &+{\sqrt{2}\over 2}\mu_{1Y}\mu_{2Y}[\hat\psi^{\dag\lambda}\hat\phi_\lambda
\hat\phi^{\dag\alpha}{1\over i\partial_-}\hat\psi_\alpha:\hat\phi^{\dag\beta}
\hat\phi_\beta:+:\hat\phi^{\dag\lambda}\hat\phi_\lambda:\hat\psi^{\dag\alpha}
{1\over i\partial_-}\hat\phi_\alpha\hat\phi^{\dag\beta}\hat\psi_\beta]\nn\\
&\ &+{\sqrt{2}\over 2}m_F\mu_{2Y}[\hat\psi^{\dag\alpha}{1\over i\partial_-}
\hat\psi_\alpha :\hat \phi^{\dag\sigma}\phi_\sigma:+
:\hat\phi^{\dag\sigma}\hat\phi_\sigma:\hat\psi^{\dag\alpha} 
{1\over i\partial_-}\hat \psi_\alpha]\nn\\
&\ & +{\sqrt{2}\over 2}\mu_{2Y}^2:\hat\phi^{\dag\sigma}\hat\phi_\sigma:
\hat\psi^{\dag\alpha}{1\over i\partial_-}\hat\psi_\alpha:\hat\phi^{\dag\beta}
\hat \phi_\beta:
\Big)
.\eea
This Hamiltonian as it stands is not normal ordered, to 
define  it properly we need to normal order the color singlet products
of bosons in the sixth term and the products of fermions in 
the last three  terms.
All these terms except one  will give some divergences which can be 
cancelled by redefinitions of $m_{B0}$ and $\lambda^2_{B0}$
in the original Hamiltonian as we will see.
One of them cannot be removed by the original terms in the 
action and we will add a counter term which cannot be put into the
original action.
This Hamiltonian could be a better two dimensional representative 
of the four dimensional Yukawa theory, since in four 
dimensions phi-four coupling is necessary to 
renormalize the Yukawa interaction.

Let us  recall the quantization of this system in the light-cone coordinates,
the Fourier mode expansions read, 
$$
\phi_{\alpha}(x^{-}) = \int a_{\alpha}(p)e^{-ipx^{-}}\frac{[dp]}{
(2|p|)^{1/2}} \, ,  \qquad \psi_{L\alpha}(x^{-})= \int \chi_{\alpha}(p)
e^{-ipx^{-}}  \frac{[dp]}{ 2^{1/4}},$$
The normalization factors are chosen to reproduce 
 the correct classical limits.
To precisely define these expansions, we assume that the momenta 
range between $(-\infty, -\epsilon_0]$ and $[\epsilon_0, \infty)$, at the 
end of our calculations we set $\epsilon_0\to 0$. This is physically
meaningful due to charge conjugation invariance, and amounts to the 
principal value prescription.
(see \cite{tolyateo, istlect} for details) 
\begin{equation}
[ \chi_{\alpha}(p) , \chi^{\dagger\beta} ( q) ]_+ = 
\delta^{\beta}_{\alpha} \delta[p - q] \, , \qquad 
[a_{\alpha}(p), a^{\dagger\beta} (q)] = {\rm sgn} (p)
\delta^{\beta}_{\alpha}\delta[p -q] \, . \end{equation}
One defines a   Fock vacuum state $|0\rangle $ by conditions,
\beq
 a_{\alpha}(p)|0\rangle = \chi_{\alpha}(p)|0\rangle =0\ \ {\rm for }\ \ p>0
 \quad   a^{\dagger\alpha}(p)|0\rangle =
 \chi^{\dagger\alpha}(p)|0\rangle =0 \  \  {\rm for}\ \  p< 0.
\eeq 
(recall that we are assuming that there is an infinitesimal hole around 
$p=0$ to be taken to zero at the end of the calculations).
The corresponding normal orderings are defined in \cite{tolyateo}.
It is useful to keep in mind the 
normal ordering rules of the bilinears,
\beq
:a^{\alpha\dag}(p)a_\beta(q):=a^{\alpha\dag}(p)a_\beta(q)-{1\over 2}
\delta^\alpha_\beta(1-\sgn(p))\delta[p-q]
.\eeq
and 
\beq
 :\chi^{\dag\alpha}(p)\chi_\beta(q):=\chi^{\dag\alpha}(p)\chi_\beta(q)
-{1\over 2}\delta^\alpha_\beta(1-\sgn(p))\delta[p-q]
.\eeq

We provide some of the  details of the reorganization of the Hamiltonian 
into normal ordered bilinears in the appendix.
We formulate the theory in terms of the color invariant bilinears 
following the idea proposed by Rajeev \cite{2dqhd}
and use our results in \cite{tolyateo}.
For the convenience of the reader we recollect some of the essential
points:
to define the large-$N_c$ limit we  introduce,  
\bea
\label{bilin}  \hat  M(p,q)&=&{2 \over  N_c}
:\chi^{\dagger \alpha}(p)\chi_\alpha(q):\cr 
  \hat  N(p,q)&=&{2 \over  N_c}:a^{\dagger \alpha}(p)a_\alpha(q):
\eea
and their odd counterparts, 
\bea 
 \hat Q(p,q)={2\over N_c}\chi^{\dagger\alpha}(p)a_\alpha(q) \, ,
\quad \hat{\bar Q}(r,s)={2 \over N_c}a^{\dagger \alpha}(r)\chi_\alpha(s)
\eea

In the large-$N_c$ limit these bilinears become 
classical variables \cite{yaffe}, and we 
postulate 
the super Poisson brackets satisfied by these
variables, which   defines the kinematics of our theory: 
\begin{eqnarray}  
 \{ M(p,q), M(r,s)\}&=&- 2i \bigl[  M(p,s)\delta[q-r] -  
  M(r,q)\delta[p-s]\nn\\ 
  &-&  \delta[p-s]\delta[q-r](\sgn(p)-\sgn(q)) \bigr] \,  \nn\\ 
   \{  N(p,q), N(r,s) \} & =&   
- 2i \bigl[  N(p,s)\sgn(q)\delta[q-r]- N(r,q)\nonumber  
\sgn(p)\delta[p-s] \nn\\
&- & \delta[q-r]\delta[p-s](\sgn(p)-\sgn(q)) \bigr] \nn\\
 \{  Q(p,q), {\bar Q}(r,s) \}_+  
&=&-2i  \bigl[  M(p,s)\sgn(q)\delta[q-r]+ N(r,q) \delta[p-s]\nn \\  
&+&\delta[p-s]\delta[q-r](1-\sgn(p)\sgn(q)) \bigr] \,  
 \nn \cr \{  M(p,q), Q(r,s)\} 
&=&-2i\delta[q-r] Q(p,s) \nn \\
\{ N(p,q), Q(r,s)\}&=&2i\delta[p-s]\sgn(p)   
 Q(r,q)\nn \cr \{  M(p,q),{\bar Q}(r,s)\} 
&=&2i\delta[p-s] {\bar  Q}(r,q)\nn \\  
\{ N(p,q),{\bar Q}(r,s)\} &=&-2i\delta[q-r]  
\sgn(q){\bar Q}(p,s)  
.\end{eqnarray}  
These classical variables  now satisfy,
\beq
    M(p,q)=M^*(q,p)\quad N(p,q)=N^*(q,p)\quad \bar Q(p,q)=Q^*(q,p)
.\eeq
There are also constraints satisfied by these variables when we  restrict
ourselves   to  the subspace of the 
color invariant states.
For our problem, {\it this is an approximation,
since there is no reason to 
expect that all the physical states are color 
singlets.} In fact we will see that there are 
scattering states of our linearized equations.
We write explicitly  the constraints for the basic variables,
\begin{eqnarray}
    (M+\epsilon)^2+Q\epsilon  Q^\dag &=&1\nn \\
     \epsilon  Q^\dag M+\epsilon  Q^\dag  \epsilon +\epsilon N 
\epsilon Q^\dag + Q^\dag &=&0\nn \\
     M Q+\epsilon Q+Q\epsilon N+Q\epsilon&=&0\nn \\
 (\epsilon N+\epsilon)^2+\epsilon Q^\dag Q&=&1\, . 
\end{eqnarray}
Above we use an operator notation, $\epsilon(p,q)=-\sgn(p)\delta[p-q]$,
and $(AB)(p,q)=\int ds A(p,s)B(s,q)$.
The phase space of the resulting restricted theory is shown to be 
a super-Grassmannian in \cite{tolyateo}, with its natural symplectic 
sturcture generalizing the results  in \cite{2dqhd}.

We can reexpress our Hamiltonian in terms of the above mentioned  
basic variables. 
After a somewhat long but 
straigthforward computation,   the large-$N_c$ limit  Hamiltonian becomes,
\beq
    H=H_0+H_Y
,\eeq
where,
\beq
 H_0={1\over 4}m_F^2\int {[dp]\over p}M(p,p)+
{1\over 4}m_{BR}^2 \int {[dp]\over |p|}
N(p,p),
\eeq
\begin{eqnarray}
H_Y&=& \int {[dpdqdsdt]\over \sqrt{|sq|}}\delta[p-q+s-t]
\Big( {1\over 16}\mu_{1Y}m_F\Big[{1\over p}+{1\over t}\Big]+
\kappa {1\over s-t}\Big)Q(p,q)\bar Q(s,t)\nn\\
&-&{\mu_{2Y}m_F\over 16}\int {[dpdqdsdt]\over \sqrt{|st|}}\delta[p-q+s-t]
\Big[ {1\over p}+{1\over q}\Big]M(p,q)N(s,t)\nn\\
&+&{1\over 64}\int {[dpdqdkdl]\over \sqrt{|pqkl|}}\lambda_R^2\delta[p-q+k-l]
N(p,q)N(k,l)\nn\\
&-&\int {[dpdqdsdtdkdl]\over \sqrt{|qskl|}}
\delta[p-q+s-t+k-l]\nn\\
&\ & \ \ \ \ \times \Big[ {\mu_{1Y}\mu_{2Y}\over 64}
\Big( {1\over k-l-t}+{1\over s-t-q}\Big)+{\mu_{1Y}^2\over 64}\Big({1\over
s-t-l}\Big)\Big]Q(p,q)\bar Q(s,t)N(k,l)\nn\\
&+&{\mu_{2Y}^2\over 64}\int {[dpdqdsdtdkdl]\over \sqrt{|pqkl|}}
\delta[p-q+k-l+s-t]{1\over t+l-k}N(p,q)N(k,l)M(s,t) \label{hamm}
.\eea
Note that we have rescaled the coupling constants as  
$\mu_{1Y} N_c\mapsto \mu_{1Y}, \mu_{2Y}N_c\mapsto \mu_{2Y}$ and 
$\lambda^2_R N_c \mapsto \lambda^2_R$.

As we discussed in the appendix, 
there are two possible renormalizations.
In the first case we allow for the 
non-local counter terms and remove the divergent parts 
finding a local Hamiltonian--thus $\kappa$ and $\lambda_{RB}$ 
are just constants.
If we only allow for the local counter terms then we find,
\bea
\kappa&=&\Big[\kappa_{R}(\mu_R)
-{\mu_{1Y}^2\over 64\pi}\ln\Big|{s-t\over \mu_R}\Big|\Big]\nn\cr
\lambda^2_{R}&=&
{\lambda^2_{RB}(\mu_R)}+{\mu_{2Y}^2\over\pi}\ln\Big|{k-l\over
\mu_R}\Big|
.\eea
We introduce a renormalization scale $\mu_R$, and assume that 
the renormalized values of the couplings, 
$\kappa_R(\mu_R), \lambda^2_{RB}(\mu_R)$  
vary with the scale $\mu_R$ such that the Hamiltonian 
does not really depend on this scale.
This means we should impose,
\begin{eqnarray}
   \kappa_R(\mu_R)&=&\kappa_R(\tilde \mu_R)
-{\mu_{1Y}^2\over 64 \pi}\ln \Big|{\mu_R\over \tilde \mu_R}\Big|\nn\\
 \lambda^2_{RB}(\mu_R)&=&\lambda_{RB}^2(\tilde \mu_R)+{\mu_{2Y}^2\over \pi}
\ln\Big|{\mu_R\over \tilde \mu_R}\Big|
.\eea
The sign of the new coupling $\kappa_R$ should not  be fixed since
it does  not exist in the original action, and 
{\it it has dimensions of mass square}.

{\it For the rest of this work we will take the simpler 
Hamiltonian, that is we  assume that all the 
renormalized couplings are ordinary numbers}.
From a more conservative point of view the local counter terms 
should be the general class of models we should investigate.
We hope to return to a more detailed analysis in the 
future.

This Hamiltonian along with the Poisson brackets and the contraint define our
model completely. As it stands this is a complicated system.
We plan to study  a variational approach to this model in a 
future work. 
We will study a linear approximation to this model in the 
next section.

\section{The linear approximation}
 
We assume that all the basic variables 
deviate from the vacuum by small amounts, therefore 
we keep everything to 
first order. 
This means the linearization of the constraint and the linearization of the
equations of motion. 
The constraint implies that 
\beq
   M(u,v)=0 \quad N(u,v)=0\quad Q(u,v)=0 \ {\rm if} \ uv>0
.\eeq
The equations of motion for $u>0, v<0$, found from 
\beq
   {\partial O(u,v)\over \partial x^+}=
\{ O(u,v), H\},
\eeq
where $O$ refers to any one of our variables, could be linearized. 
Let us write these linearized equations of motion for all the variables,
\bea
  {\partial M(u,v)\over \partial x^+}=i{m_F^2\over 2}\Big[{1\over u}-
{1\over v}\Big]M(u,v)-
i{\mu_{2Y}m_F\over 4}\Big[ {1\over u}+{1\over v}\Big]
\int {N(s, s-(u-v))\over \sqrt{|s((u-v)-s)|}}[ds]
,\eea
\bea
  {\partial N(k,l)\over \partial x^+}&=&i{m_{BR}^2\over 2}\Big[{1\over k}
-{1\over l}\Big]N(k,l)
-i{\mu_{2Y}m_F\over 4\sqrt{|kl|}}\int [ds]\Big[{1\over s}+
{1\over s-(k-l)}\Big]M(s, s-(k-l))\nn\\
&+&{i\over 8}\lambda_{R}^2\int {[ds]\over \sqrt{|skl((k-l)-s)|}}N(s, s-(k-l))
,\eea
and 
\bea
{\partial Q(u,v)\over \partial x^+}&=&{i\over 2}\Big[{m_F^2\over u}-
{m_{RB}^2\over v}\Big]Q(u,v)-\kappa {4i\over u-v}\int 
{Q(p, p-(u-v))\over \sqrt{|v(p-(u-v))|}}[dp]\nn\\
&+& i{\mu_{1Y}m_F\over 4}\int {[dp]\over \sqrt{|(p-(u-v))v|}}
\Big[{1\over p}+{1\over u}\Big]Q(p, p-(u-v))
.\eea
The equation of motion for $\bar Q$ can be found by complex conjugation 
and does not carry new information.
We note that the equations of motion for $M$ and $N$ are coupled,
but the equations of motion for $Q$ within the linear approximation 
is decoupled from the rest.

So we will start with this one and make an ansatz as in \cite{2dqhd, istlect}.
Let us assume that  the solution can be written as  
$Q(u,v;x^+)=c_Q(x)e^{iP_+x^+}$, where 
$x=u/P_-$, $P_-=u-v$ and define an invariant mass
$\Lambda_Q^2=2P_-P_+$. (Strictly speaking 
we could take  the solution of the form
$c_Q(x)f_Q(P_-)e^{iP_+x^+}$, the arbitrary function
$f_Q$ factors out in the equations). Then  we find,
\bea
   \Lambda_Q^2 c_Q(x)&=&\Big[ {m_F^2\over x} +{m_{BR}^2\over 1-x}\Big]c_Q(x)
-8\kappa\int_0^1 {dy\over \sqrt{(1-y)(1-x)}}c_Q(y)\nn\cr
&+& {1\over 4} \mu_{1Y}m_F\int_0^1 {dy\over \sqrt{(1-y)(1-x)}}\Big( {1\over x}
+{1\over y}\Big) c_Q(y)
.\eea 
This innocent looking equation actually requires a renormalization,
as we will see shortly.

Let
\beq
    \int_0^1 {dy\over \sqrt{(1-y)}} c_Q(y)=A \quad 
\int_0^1 {dy\over y\sqrt{(1-y)}}c_Q(y)=B \label{cond}
.\eeq
Then solve for $c_Q(x)$,
\beq
   c_Q(x)=\sqrt{(1-x)}{(aA+bB)x+bA\over
\Lambda_Q^2x(1-x)-m_F^2(1-x)-m_B^2x}
,\eeq
where 
\beq  
    a=-8\kappa,
\quad \quad  b={1\over 4}\mu_{1Y}m_F
.\eeq
A straigthforward solution will actually produce a divergence,
the integration defining $B$ is divergent. 
To find a finite result we need a renormalization prescription.
Let us assume that the phase is given by
$e^{i(P_++\delta P_+(\epsilon_0))x^+}$, where 
$\delta P_+(\epsilon_0)$ denotes a divergent phase 
of the solution that we remove from the equations, and 
$\epsilon_0$ denotes a low momentum  cut-off.
The time derivative will drop a factor of $\delta P_+(\epsilon_0)$,
and multiplying by $2P_-$ we denote it as $\delta \Lambda_Q(\epsilon_0)$ and 
rewrite the same equation as,
\beq
 \Big[ \Lambda_Q^2 +\delta \Lambda_Q(\epsilon_0)-{m_F^2\over x}
-{m_{BR}^2\over 1-x}\Big]c_Q(x)=
{aA\over \sqrt{1-x}}+ {bA\over x\sqrt{1-x}}+{bB(\epsilon_0)\over \sqrt{1-x}}
.\eeq   
Since the divergent part comes from the $B$ term we expect that 
 $\delta \Lambda_Q(\epsilon_0)c_Q(x)$ can be 
taken as a counterm on the other side of the equality  with 
the leading form $ -{\alpha_c(\epsilon_0)\over \sqrt{1-x}}$.
The unknown function now is given by the same 
formula  with a shifted coefficient of $x$,
\beq
   c_Q(x)=\sqrt{(1-x)}{(aA+bB(\epsilon_0)-\alpha_c(\epsilon_0))x+bA\over
\Lambda_Q^2x(1-x)-m_F^2(1-x)-m_B^2x}
.\eeq
Let us insert this back into (\ref{cond}) and find the constants
$A, B(\epsilon_0)$.
After some algebra, we reach,
\beq
A=-F_2 {aA+bB(\epsilon_0)-\alpha_c(\epsilon_0)\over 2\Lambda_Q^2}
 -\Big[ \Big({aA+bB(\epsilon_0)-\alpha_c(\epsilon_0)\over 
2\Lambda_Q^2}\Big)(\Lambda_Q^2-m_{RB}^2+m_F^2)+bA\Big]F_1(\Lambda_Q)
,\eeq
\beq
B(\epsilon_0)={bA\over m_F^2}\ln(\epsilon_0)+
{bA\over 2m_F^2}F_2-\Big[ {bA\over 2m_F^2}(\Lambda_Q^2-m_{RB}^2+m_F^2)
+(aA+bB(\epsilon_0)-\alpha_c(\epsilon_0))\Big]F_1(\Lambda_Q) 
,\eeq
where,
\bea
  F_2&=&\int_0^1 dx { 2\Lambda_Q^2x-(\Lambda_Q^2-m_{RB}^2+m_F^2)\over
\Lambda_Q^2 x^2-(\Lambda_Q^2-m_{RB}^2+m_F^2)x+m_F^2}=
\ln\Big[{m_{RB}^2\over m_F^2}\Big],\cr
F_1(\Lambda_Q)&=&\int_0^1  { dx \over
\Lambda_Q^2 x^2-(\Lambda_Q^2-m_{RB}^2+m_F^2)x+m_F^2}
.\eea
If we are looking for a bound state of a  boson and a fermion,
this requires,
\beq 
|m_{RB}-m_F|<\Lambda_Q<m_{RB}+m_F,
\eeq
 then the  last integral gives,
\beq
  F_1(\Lambda_Q)={2\over u}\Big(\arctan\Big[{\Lambda_Q^2+m_{BR}^2-
m_F^2\over u}\Big]-\arctan\Big[{m_{BR}^2-
m_F^2-\Lambda_Q^2\over u}\Big]\Big),
\eeq
where,
\beq
   u=\sqrt{(\Lambda_Q^2-(m_{RB}-m_F)^2)((m_{RB}+m_F)^2-\Lambda_Q^2)}
.\eeq
Let us impose the two conditions,
\beq
   B_*=B(\epsilon_0)-{bA\over m_F^2}\ln(\epsilon_0), \quad
 bB(\epsilon_0)-\alpha_c(\epsilon_0)=bB_*
,\eeq
then we see that if we set 
\beq
\alpha_c(\epsilon_0)={b^2 A\over m_F^2}\ln(\epsilon_0)
,\eeq
we can take $\epsilon_0\to 0^+$ 
limit, and keep $B_*, A$ finite.
The renormalized equations become,
\beq
A=-F_2 {aA+bB_*\over 2\Lambda_Q^2}
 -\Big[ \Big({aA+bB_*\over 
2\Lambda_Q^2}\Big)(\Lambda_Q^2-m_{RB}^2+m_F^2)+bA\Big]F_1(\Lambda_Q)
,\eeq
\beq
B_*=
{bA\over 2m_F^2}F_2-\Big[ {bA\over 2m_F^2}(\Lambda_Q^2-m_{RB}^2+m_F^2)
+(aA+bB_*)\Big]F_1(\Lambda_Q) 
.\eeq
If we solve for the ratio  $A/B_*$, after some algebra, this 
gives us a consistency  condition for 
the excitation energy $\Lambda_Q^2$, 
\bea
&\  &\Big[ F_2-s(\Lambda_Q)F_1(\Lambda_Q)-{2am_F^2\over b}F_1(\Lambda_Q)\Big]
\Big[ F_2+s(\Lambda_Q)F_1(\Lambda_Q)\Big]=\cr
&\ &\ \ \ \ \ \ \ \ \ \quad \quad \quad 
-\Big[ 1+bF_1(\Lambda_Q)\Big]\Big[ F_2 +
(s(\Lambda_Q)+{2b\Lambda_Q^2\over a})F_1(\Lambda_Q)+{2\Lambda_Q^2\over a}\Big]
{2m_F^2a\over b^2}
,\eea
or equivalently,
\bea
&\  &\!\!\!\!\!\!\!\!\Big[ \ln\Big|{m_{RB}^2\over m_F^2}\Big|
-s(\Lambda_Q)F_1(\Lambda_Q)+{64\kappa m_F\over \mu_{1Y}}F_1(\Lambda_Q)\Big]
\Big[\ln\Big|{m_{RB}^2\over m_F^2}\Big| +s(\Lambda_Q)F_1(\Lambda_Q)\Big]=\cr
&\ &\ \ \ \ \ 
\Big[ 1+{1\over 4}\mu_{1Y}m_FF_1(\Lambda_Q)\Big]
\Big[ \ln\Big|{m_{RB}^2\over m_F^2}\Big| +
(s(\Lambda_Q)-{\mu_{1Y}m_F\Lambda_Q^2\over 16\kappa})F_1(\Lambda_Q)+
{\Lambda_Q^2\over 4\kappa}\Big]
{128 \kappa\over \mu_{1Y}^2}
,\eea
where $s(\Lambda_Q)=\Lambda_Q^2-m_{RB}^2+m_F^2$.
This equation is written in terms of dimensionless ratios of the variables,
and it should be investigated numerically under the conditions we have 
stated before for $\Lambda_Q$.
Instead of numerically solving these equations we will look 
at one  extreme case,
when $\Lambda_Q\approx m_{BR}+m_F$ (weak coupling). 
(It is interesting to investigate 
the opposite limit $\Lambda_Q\approx |m_F-m_{BR}|$, but 
the result is not so simple to interpret).
It is better to use a different variable to 
study such limiting  cases,
we define $\Delta$, via 
$\Lambda_Q^2=m_{BR}^2+m_F^2+2m_{BR}m_F\Delta$.
Note that we have $-1<\Delta<1$.
Our function $F_1(\Lambda_Q)$ becomes,
\beq
    F_1(\Delta)={1\over m_F m_{BR}} {1\over \sqrt{1-\Delta^2}}
\Big(\arctan\Big[{\omega+\Delta\over \sqrt{1-\Delta^2}}\Big]
+\arctan\Big[ {1/\omega +\Delta\over \sqrt{1-\Delta^2}}\Big]\Big)
.\eeq
Here $\omega=m_{BR}/m_F$ and if we take 
$\Delta\to 1^-$, that means $\Lambda_Q\to (m_{BR}+m_F)^-$ 
and $\Delta\to -1^+$ corresponds to $\Lambda_Q\to |m_F-m_{BR}|^+$.
There is nothing subtle about the first limit,
keeping everything to first order  gives us,
\beq
 \Delta=1-\Bigg[1-\Big(2-{\mu_{1Y}\over \kappa}
(m_F+m_{BR})\Big)^{-1}
{\mu_{1Y}\over m_F+m_{BR}}
\ln\Big|{m_{BR}^2\over m_F^2}\Big|\Bigg]^{-2}{\pi^2\mu_{1Y}^2\over
m_B^2}
.\eeq
We assumed all the way $\Delta\approx 1$, 
this  could be consistent if for example we 
choose the coupling constant $\mu_{1Y}$ such that $\mu_{1Y}<<m_B$ when 
the ratio  $m_F/m_{BR}$ is not too large and if $\kappa<0$.
There are other possibilities, but this simple one shows 
that there are solutions with the expected behaviour.
If we assume $\Delta\to -1^+$,
the function $F_1(\Delta)\to 1/m_Fm_{BR}$, and the 
calculations are more complex.
If we set $\Delta=-1+\delta^2$, we have 
$F_1(-1+\delta^2)\approx {1\over m_Fm_{BR}}(1-{3\over 2}\delta^2)$.
This can be used to study the opposite limit, but 
due to its algebraic comlexity we leave it out, and only state 
that for various possible   cases to be consistent, we find 
that  $m_F>>m_{BR}$ is a necessary condition.

Following the same strategy,  we  look for  stationary solutions
for the coupled equations:  we   
start with  the following ansatz for $M,N$:
$M(u,v)=\xi_M(x)e^{iP_+x^+}$ and $N(k,l)=\xi_N(y)e^{iP_+x^+}$,
where $x=u/P_M$, $P_M=u-v$ and similarly for 
$N$. 
If we now substitute these 
into the coupled equations of motion we see that  
the oscillations in time 
cancel out (since we select the same $P_+$) and we end up with,
\beq
\Lambda_M^2\xi_M(x)=m_F^2\Big[ {1\over x}+{1\over 1-x}\Big]\xi_M(x)
-{\mu_{2Y}m_F\over 4\pi}\Big[ {1\over x}-{1\over 1-x}\Big]\int_0^1
{\xi_N(y)\over \sqrt{y(1-y)}}dy
,\eeq
\bea
\Lambda^2_N\xi_N(x)&=&m^2_{BR}\Big[{1\over x}+{1\over 1-x}\Big]\xi_N(x)
-{\mu_{2Y}m_F\over 4\pi\sqrt{(1-x)x}}\int_0^1 dy\Big({1\over y}-{1\over 1-y}
\Big)\xi_M(y)\nn\\
&+&{\lambda_R^2 \over 8\pi}\int_0^1 {dy\over \sqrt{x(1-x)y(1-y)}}\xi_N(y),
\eea
where we set $\Lambda_M^2=2P_MP_+$, and $\Lambda_N^2=2P_NP_+$ 
for the invariant masses of the excitations of $M$ and $N$ respectively.
We notice that the desired decoupling of the total momentum variables would 
not have 
 happened  in the above equations if we had used 
the more general Hamiltonians.

Before we plunge into the standart way to solve these equations 
we will talk about an interesting possibility,
if  we  admit that the distributional solutions are also 
proper solutions of these equations.
 
Let us write the equation for $M$, 
if we call
\beq
    D=\int_0^1 dy {\xi_N(y)\over \sqrt{y(1-y)}},
\eeq
then we can rewrite it as,
\beq
\Lambda_M^2\xi_M(x)=m_F^2\Big[{1-dD\over x}+{1+dD\over 1-x}\Big]\xi_M(x)
, \quad  {\rm where} \  d={\mu_{2Y}\over 4\pi m_F}
.\eeq
For the solution we have in mind we need to impose 
\beq
     dD<1
,\eeq
otherwise the energy will be unbounded from below.
Let us assume that the last two terms in the $\xi_N$ equation  cancel 
against each other. This condition implies that 
\beq
{\mu_{2Y}m_F\over 4\pi}\int_0^1 dy\Big({1\over y}-{1\over 1-y}
\Big)\xi_M(y)
={\lambda_R^2\over 8\pi}\int_0^1 {dy\over \sqrt{y(1-y)}}\xi_N(y),
\eeq
or equivalently,
\beq
{\mu_{2Y}m_F\over 4\pi}\int_0^1 dy\Big({1\over y}-{1\over 1-y}
\Big)\xi_M(y)
={\lambda_R^2\over 8\pi}D.
\eeq
Thus we have to consistently choose everything to 
satisfy these conditions.
The equation for $M$ and $N$  can be solved by  
using  
$\xi_M(x)=\delta(x-x_F)$ and $\xi_N(x)=\delta(x-x_B)$.
What should we take as $x_F$ and $x_B$?
One way is to minimize the excitation energy for fermions, 
and then fix the bosonic parameter to have the cancellation.
The eigenvalue for $M$ becomes, after the minimizing choice is made,
\beq
\Lambda_M^2=m_F^2\Big[ (1+dD)^{1/2}+(1-dD)^{1/2}\Big]^2<4m_F^2
.\eeq
The last inequality is interesting since it implies that the
fermions  actually form a bound state at this energy.
Now the condition we should have for 
the cancellation reads,
\beq
   \Big( {1+dD\over 1-dD}\Big)^{1/2}-\Big({1-dD\over 1+dD}\Big)^{1/2}=
fD,
\eeq
where 
\beq
f={\lambda_R^2 \over 2 \mu_{2Y}m_F} 
.\eeq
From here we can solve for $D$,
\beq
   D^2={1\over d^2}-{4\over f^2}
.\eeq
We should have $D^2>0$, this puts  
a  condition on  our couplings and the fermion mass.
But there is a stronger condition,
once we have the solution for the 
value of $D$, we can find the parameter 
$x_B$ to choose for the bosons from 
the definition of $D$,
\beq
     D=\int_0^1 {dy\over \sqrt{y(1-y)}}\xi_N(y)={1\over \sqrt{x_B(1-x_B)}}.
\eeq
It is possible to find $x_B$ if $D\geq 2$, since the minimum
of the function on the right is $2$,
thus we need 
$D^2\geq 4$.
This implies a condition on our couplings,
\beq
   {1\over \pi}\Big({\mu_{2Y}^2\over m_F^2}\Big)
\Big[1-{\mu_{2Y}^2\over 4\pi^2 m_F^2}\Big]^{-1/2}<
{\lambda_{RB}^2\over m_F^2}
,\eeq
where we used dimensionless variables to express this 
inequality.

We have the other condition about 
$D$, which says $dD<1$.
This is actually satisfied by our solution, so we need 
\beq
2\leq D <1/d.
\eeq
From these we have a condition on the strength of the 
Yukawa coupling  constant, 
\beq
\mu_{2Y}<2\pi m_F
.\eeq

Now we can go back and find the actual value of the fermion bound state
and the mass of the boson pair.
The boson pair mass is simply given by
\beq
  \Lambda_N^2={m_{RB}^2\over x_B(1-x_B)}=m_{RB}^2D^2=m_{RB}^2
\Big[ {1\over d^2}-{4\over f^2}\Big]=16m_{RB}^2\Big[{\pi^2m_F^2\over 
\mu_{2Y}^2}-{\mu_{2Y}^2m_F^2\over \lambda_R^4}\Big]>4m_{RB}^2.
\eeq
Similarly we have for the fermion pair,
\bea
    \Lambda_M^2&=&m_F^2\Big[\Big(1+[1-{4d^2\over f^2}]^{1/2}
\Big)^{1/2}+\Big(1-[1-{4d^2\over f^2}]^{1/2}
\Big)^{1/2}\Big]^2\nn\cr
&=&m_F^2
\Big[\Big(1+[1-{\mu_{2Y}^4\over \pi^2\lambda_R^4}]^{1/2}
\Big)^{1/2}+\Big(1-[1-{\mu_{2Y}^4\over \pi^2\lambda_R^4}]^{1/2}
\Big)^{1/2}\Big]^2
.\eea

We note that the above  solution is 
quite interesting:   we assume 
that the relative strength of the Yukawa coupling is small,
i.e. $\mu_{2Y}<<\lambda_R$,
and expand the square roots,
\beq
 \Lambda_M^2\approx 2m_F^2(1+{2d^2\over f^2})=2m_F^2
\Big[ 1+{1\over \pi} {\mu_{2Y}^4\over \lambda_R^4}\Big]
.\eeq
This is not the result one should expect from a perturbative 
point of view.
 
Furthermore  the boson pair becomes  in this  approximation, 
\beq
   \Lambda_N^2\approx 16\pi^2\Big({\mu_{2Y}\over m_F}\Big)^{-2}m_{BR}^2>
16m_{RB}^2
.\eeq
If we further assume that $\mu_{2Y}<<m_F$ this implies that the 
boson pair mass becomes very large.

Actually, there is a whole range of solutions with 
$\xi_M(x)=\delta(x-x_F)$ and $\xi_N(x)=\delta(x-x_B)$.
We are free to choose one of them, say $x_F$ then the other one 
will be determined by the same consistency relation as above.
We see that the boson pair excitation will always be
bigger than $2m_{BR}$, since the minimum is given by 
$x_B=1/2$. For the fermion pair we choose 
the consistent $x_F$'s such that the mass is less than 
the two mass threshold. 
Let us briefly present our findings using the 
same notation as above.

Let us search for a solution of 
the equation
\beq
m_F^2\Big[ {1-dD\over x}+{1+dD\over 1-x}\Big]\xi_M(x)=\Lambda_M^2\xi_M(x)
\quad {\rm with} \quad  {\Lambda_M^2 < 4m_F^2}
.\eeq
We assume again that $dD<1$, then the equality  is satisfied  if we set
$\xi_M(x)=\delta(x-x_F)$, $x_F=1/2(1-\alpha dD)$ for 
$0<\alpha<1$.
We require the same delta function solution for  $\xi_N$, this means 
we should solve for the equation,
\beq
    {1\over x_F}-{1\over 1-x_F}=fD
.\eeq
If we solve for $D$ now,
we find 
\beq
     D^2=\Big[ 1-{4\alpha d\over f }\Big]{1\over \alpha^2 d^2}
.\eeq  
This implies that the first factor in the big parathesis should be 
positive. This is true if 
$\lambda_R^2>{2\over \pi} \alpha\mu_{2Y}^2$.
Again to have a solution for $x_B$ we need $4\leq D^2$,
this means 
\beq
  4\alpha^2 d^2+{4\alpha d\over f}-1\geq 0
,\eeq
which  can be 
satisfied if the quadratic eqaution for $d$ has
real roots and we choose $d$ in between--assuming alpha is 
chosen.
This implies for $\lambda_R$ an inequality,
\beq
     {2\alpha\over \pi}\Big({\mu_{2Y}\over m_F}\Big)^2
\Big[1-\Big({\alpha\mu_{2Y}\over 2\pi m_F}\Big)^2\Big]^{-1}< {\lambda_R^2
\over m_F^2}
.\eeq
This means that we should choose $\lambda_R$ above a certain value, and this
condition is stronger than the first one we found above.
A uniform bound for various values of $\alpha$ can  be chosen,
\beq
     {2\over \pi} \Big({\mu_{2Y}\over m_F}\Big)^2 \Big[ 1-
\Big({\mu_{2Y}\over 2\pi m_F}\Big)^2\Big]^{-1}<{\lambda_R^2\over m_F^2}. 
\eeq
Incidentally this requires $\mu_{2Y}<2\pi m_F$.
We still have $dD<1$ to satisfy.
If we simply use this condition assuming $\alpha$ as given we arrive at 
the positivity of a quadratic expression in $\alpha$:
\beq 
   \alpha^2+{2\alpha\over \pi}{\mu_{2Y}^2\over \lambda_R^2}-1>0
.\eeq 
Notice that the range of allowed $\alpha$ will be bigger if we take 
$\mu_{2Y}/\lambda_R$ ratio as large as possible.
If we use the uniform lower bound  for $\lambda_R$ we will find the 
largest region, and if we denote the deviation from this value by a
 multiplicative factor $k>1$, we can insert this ratio into the 
quadratic expression,
\beq
  \alpha^2+{1\over k}\Big[1-\Big({\mu_{2Y}\over 2\pi m_F}\Big)^2\Big]\alpha-1>0
.\eeq 
The positivity is guaranteed if we choose $\alpha$ outside 
of the region between the two roots.
The choice consistent with $0<\alpha<1$ gives us,
\beq
  {1\over 2k}\Bigg( \sqrt{\Big[1-{\mu_{2Y}^2\over 4\pi^2m_F^2}\Big]^2+4k^2}
-\Big[1-{\mu_{2Y}^2\over 4\pi^2m_F^2}\Big]\Bigg)<\alpha<1
.\eeq
The lower bound is a decreasing function of $k$, so the stronger 
relative values of  $\lambda_R$ will have a smaller domain of 
$\alpha$'s. Since for these choices we have a continuous range 
of $\alpha$'s, the spectrum of the problem is rather different.
Such distributional solutions are not eigenvalues but they typically
refer to the continuous part of the spectrum.
In some sense these are still scattering states.  
This means we cannot use the free parts  of the 
original Hamiltonian to study the scattering theory 
below the two mass thresholds. Above these values 
we will see that the scattering theory can be studied by conventional 
methods. Perhaps below this we need to use the minimum value 
of the spectrum to define new effective pair mass for the fermionic 
sector. We are not able to resolve this issue at the moment.

If we now go back to the standart approach,
again as in the case of boson-fermion pair, we will need
to renormalize our equations, by assuming a divergent common 
phase $\delta P_+(\epsilon_0)$,
$2i\xi_{M}e^{i(\delta P_+(\epsilon_0)+P_+)x^+}$, and 
similarly for $N$.
The derivative will bring terms of the form
$\delta \Lambda_N(\epsilon_0)=2P_N\delta P_+(\epsilon_0)$
and $\delta \Lambda_M(\epsilon_0)=2P_M\delta P_+(\epsilon_0)$.
An inspection of the resulting equations show that,
we have the leading behaviour  
\beq
\delta \Lambda_M(\epsilon_0)\xi_M(x)
\sim{\alpha_c(\epsilon_0)(1-2x)\over x(1-x)},
\quad \delta \Lambda_N(\epsilon_0)\xi_N(x)\sim
{\beta_c(\epsilon_0)\over \sqrt{x(1-x)}}
.\eeq
In general, $\beta_c(\epsilon_0)=\sigma\alpha_c(\epsilon_0)$,
as we will see the precise value of $\sigma$ is not 
important.
Now we can solve the unknown functions,
\bea
 \xi_M(x)&=&{(1-2x)(aA(\epsilon_0)-\alpha_c(\epsilon_0))\over
\Lambda_M^2x(1-x)-m_F^2}\cr
\xi_N(x)&=& \sqrt{x(1-x)}{ 2aB(\epsilon_0)+bA(\epsilon_0)-
\sigma\alpha_c(\epsilon_0)\over \Lambda_N^2x(1-x)-m_{BR}^2}
,\eea
where 
\beq
   A(\epsilon_0)=\int_{\epsilon_0}^1 dy{\xi_N(y)\over \sqrt{y(1-y)}}
\quad 
B(\epsilon_0)=\int_{\epsilon_0}^1 dy{\xi_M(y)\over y}
,\eeq
and 
\beq
   a=-{\mu_{2Y}m_F\over 4\pi}\quad b={\lambda_R^2\over 8\pi}
.\eeq

If we introduce
\beq
    F(\Lambda, m)={\cal P}\int_0^1 {dy\over \Lambda^2y(1-y)-m^2},
\eeq
we find from the defining conditions of $A(\epsilon_0), B(\epsilon_0)$,
\bea
B(\epsilon_0)&=& (aA(\epsilon_0)-\alpha_c(\epsilon_0)){1\over m_F^2}
\ln \epsilon_0+(aA(\epsilon_0)-\alpha_c(\epsilon_0))\Big({\Lambda_M^2\over 2m_F^2}-2\Big)F(\Lambda_M,m_F)\cr
A(\epsilon_0)&=&\Big[ 2aB(\epsilon_0)+
bA(\epsilon_0)-\sigma\alpha_c(\epsilon_0)
\Big]F(\Lambda_N^2,m_{RB})
.\eea
We now define 
$aA_*=aA(\epsilon_0)-\alpha_c(\epsilon_0)$, and 
$B_*=B(\epsilon_0)-{a\over m_F^2}A_*\ln\epsilon_0$,
and insert these back into our equations,
\bea
B_*&=& A_*a\Big({\Lambda_M^2\over 2m_F^2}-2\Big)F(\Lambda_M,m_F)\cr
A_*&=&\Big[ 2aB_*+bA_*+{2a^2\over m_F^2}A_*\ln \epsilon_0+
[{b\over a}-\sigma-{1\over aF(\Lambda_N, m_{RB})}]\alpha_c(\epsilon_0)
\Big]F(\Lambda_N,m_{RB})
.\eea
If we set
\beq
   \alpha_c(\epsilon_0)=-\Big[{b\over a}-\sigma-
{1\over aF(\Lambda_N, m_{RB})}\Big]^{-1}
{2a^2\over m_F^2} A_*\ln \epsilon_0
,\eeq
we can   take   $\epsilon_0\to 0^+$ limit
while keeping $A_*, B_*$ finite.
These will be our renormalized equations,
\beq
B_*=aA_*\Big( {\Lambda_M^2\over 2m_F^2}-2\Big)F(\Lambda_M,m_F)
\quad 
A_*=[2aB_*+bA_*]F(\Lambda_N, m_{BR})
.\eeq
Incidentally we note that this physical prescription 
implies that the proper way  we should define these 
integral equations is to use the Hadamard finite value
(see the similar issue in \cite{istlect} and 
\cite{gracia}  for a recent discussion of the renormalization and
distribution theory).

Let us assume that we are looking for bound state solutions,
the principal value integral in $F(\Lambda,m)$ then 
becomes an ordinary integral. 
Now we have two different expression for the 
ratio  
$B_*/A_*$, which give us the desired eigenvalues when we 
require square integrable solutions.
If we assume that both of the eigenvalues are bound states we find,
\beq
   \Bigg[{\mu_{2Y}^2\over 4\pi^2}{\sqrt{\Lambda_M^2-4m_F^2}\over 
\Lambda_M}\arctan\Big[{\Lambda_M\over \sqrt{4m_F^2-\Lambda_M^2}}\Big]
+{\Lambda_R^2\over 8\pi}\Bigg]{2\over \Lambda_N\sqrt{4m_{BR}^2-\Lambda_N^2}}
\arctan \Big[{\Lambda_N\over \sqrt{4m_{BR}^2-\Lambda_N^2}}\Big]=-1
.\eeq
Since the left hand-side is positive and the right one is 
negative this has no solution! {\it if 
we demand both boson-boson and fermion-fermion pairs 
to form bound states, there is no solution}.
But we should not be alarmed, what we can do is to demand a
resonance for the bosonic sector, then,
\beq
   \Bigg[{\mu_{2Y}^2\over 4\pi^2}{\sqrt{\Lambda_M^2-4m_F^2}\over 
\Lambda_M}\arctan\Big[{\Lambda_M\over \sqrt{4m_F^2-\Lambda_M^2}}\Big]
+{\Lambda_R^2\over 8\pi}\Bigg]{2\over \Lambda_N\sqrt{\Lambda_N^2-4m_{BR}^2}}
\ln\Big|{\Lambda_N+\sqrt{\Lambda_N^2-4m_{BR}^2}\over\Lambda_N-
\sqrt{\Lambda_N^2-4m_{BR}^2}}\Big|=1
,\eeq
and this has  solutions in general.
The resonance case seems to be special to 1+1 dimensions, 
the other possibility is to require bosons to have 
scattering states and fermions to be bound.
This  can be studied  by analyzing the
pole structure of the analytic continuation of the 
scattering amplitudes that will be worked out
below. Due to the algebraic complexity of the 
resulting formulae, 
we will not be able to answer it in this work.

We study the scattering  states when we are 
beyond the two mass threshold. 
Let us make a digression  for the moment and study a simpler problem,
the lambda-phi-four coupling.
It is easy  to see from our equtions for $N(u,v)$ that the same 
ansatz for the solution leads to 
\beq
   \Lambda^2 \xi(x)={m^2\over x(1-x)}\xi(x)+{\lambda_B^2\over 8\pi}
{1\over \sqrt{x(1-x)}}\int_0^1 {\xi(y)\over \sqrt{y(1-y)}}
.\eeq
We assume that the operator on the right is acting  on 
$L^2([0,1])$, with vanishing at the end points boundary conditions.
The free part, 
\beq
 H_0=  {m^2\over x(1-x)}
\eeq
is an unbounded operator with a continuous spectrum 
 $[4m^2, \infty)$.
This is easy to understand by studying a particle and antiparticle 
in the center of momentum frame. 

If the added term is not a ``too strong''  perturbation 
then the absolutely continuous part of the 
spectrum of the full operator on the 
right is the same as the spectrum of the 
free part and we can study the scattering 
states using the free part (there are various conditions 
we can state so that ``too strong'' becomes a 
precise statement, we recommend  \cite{reedsimon} for a thorough
mathematical discussion of these issues, 
and \cite{newton} with more physical emphasis).
The type of problem we study is analyzed in a recent valuable 
book by Albeverio and Kurasov \cite{albeverio}.
The interaction term is called a rank one perturbation.
The Hamiltonian,
\beq
  H={m^2\over x(1-x)}+{\lambda_B^2\over 8\pi}{1\over \sqrt{x(1-x)}}\int_0^1
dy {1\over \sqrt{y(1-y)}}
,\eeq
where everything  acts on functions in  $L^2([0,1])$,
 can be written as
\beq
    H=H_0+{\lambda_B^2\over 8\pi}|f><f|
,\eeq
with $<x|f>=f(x)={1\over \sqrt{x(1-x)}}$.
If such a perturbation is relatively form bounded then 
the scattering states are given by the scattering states 
of the free part.
To verify this condition it is enough to  show that 
the added term satisfies 
\beq
    ||f||_{-1}=||{1\over (|H_0|+1)^{1/2}}f||<\infty
,\eeq
where $||.||$ denotes the usual $L^2$ norm.
It is now simple to check that,
\beq
  ||f||_{-1}^2<   \int_0^1 dx
\Big({m^2\over x(1-x)}\Big)^{-1}\Big({1\over \sqrt{x(1-x)}}\Big)^2<\infty
.\eeq
In fact in the above problem we can find the
resolvent of  our integral operator 
(see \cite{albeverio}):
\begin{eqnarray}
(R_H(Z)f)(x)&=&[(H-Z)^{-1}f](x)\nn\cr
  &=&\Big({m^2\over x(1-x)}-Z\Big)^{-1}f(x)
+{\lambda_B^2\over 8\pi}\tilde A(Z){\sqrt{x(1-x)}\over Zx(1-x)-m^2}\int_0^1
{\sqrt{y(1-y)}dyf(y)\over Zy(1-y)-m^2}
,\end{eqnarray}
for $Z$ outside of the spectrum (and complex in general), and here
we use the analytic continuation of 
 $\tilde A(\lambda)$ to complex numbers and its explicit form 
is given below.
The knowledge of the resolvent gives everything about the 
operator, for example we can find the 
spectral density function $\rho(\Lambda)$ (which heuristically 
corresponds to the ``eigenfunction'' expansion) by using the 
well-known Stone's identity,
\beq
     \rho(\Lambda)={1\over \pi i}
\lim_{\epsilon_0\to 0^+} \Bigg( {1\over H-\Lambda+i\epsilon_0}-
{1\over H-\Lambda -i\epsilon_0}\Bigg)
.\eeq 
In \cite{albeverio} the scattering theory of finite rank perturbations
has been worked out by using rigorous methods. 
We will only content with the result that the scattering theory
makes sense {\it beyond the bound state 
thresholds for both particles}, and we can find the resolvents explicitly. 
For simplicity of our presentation  we  study the 
scattering theory by the standart methods  in physics, and 
only find the wave operators.
One can see that for a given value of $\Lambda^2={m^2\over \lambda(1-\lambda)}$
we have two roots,
\beq
      \lambda_\pm ={1\over 2}\pm \Big[ {1\over 4}-
{m^2\over \Lambda^2}\Big]^{1/2}
.\eeq
In physics we typically think of two particles approaching and 
then scattering off to infinity.
An inspection of the kinematics of a particle and an antiparticle 
pair in the center of momentum frame reveals that 
 $\lambda_+$ corresponds to the particle moving in the positive
$x^1$ direction (which we may take as ``incoming'' states),
and $\lambda_-$ corresponds to the particle moving in the
negative $x^1$ direction. 
From a physical point of view, 
the scattering data should give us the information about transmission
and reflection of the pair.
An equivalent description would be to find the 
wave operator, $\Omega$ which maps 
(in general the projection to the absolutely continuous part of the spectrum
of the original operator)  the Hilbert space 
 to the scattering states (if we take the absolutely
continuous part of the spectrum and use the spectral projections 
corresponding to these values) of the interacting Hamiltonian:
\beq
    \xi=\Omega f
.\eeq
(In physics one typically uses $\Omega_+$ which takes the 
wave functions at time zero and evolve them to positive infinity,
this requires the $+i\epsilon$ prescriptions in the integrals,
we will see that for our problem it is more suitable to 
define the principal value one, this is why we use $\Omega$).
    
To find the scattering amplitudes we rewrite the 
eigenvalue equation in 
Lippmann-Schwinger form,
\beq
  \xi_\lambda(x)=\delta(x-\lambda)+{\lambda^2_B\over 8\pi}\Big[\Lambda^2-
{m^2\over x(1-x)}\Big]^{-1}{1\over \sqrt{x(1-x)}} 
\int_0^1 dy {\xi_\lambda(y)\over \sqrt{y(1-y)}}
.\eeq
We formally use the eigenvalue $\delta_\lambda(x)=\delta(x-\lambda)$ of the 
free Hamiltonian.
We can now solve for 
$A(\lambda)=\int_0^1 {\xi_\lambda(y)\over \sqrt{y(1-y)}}$,
\beq
A(\lambda)=\Big[1+{\lambda_B^2\over 4\pi \Lambda \sqrt{(\Lambda^2-4m^2)}}
\ln\Big|{\Lambda-\sqrt{\Lambda^2-4m^2}\over 
\Lambda+\sqrt{\Lambda^2-4m^2}}\Big|\Big]^{-1}
{1\over \sqrt{\lambda(1-\lambda)}}
,\eeq
where $\Lambda^2={m^2\over \lambda(1-\lambda)}$.
Thus the wave operator acting on the 
(formal) eigenfunctions of the free Hamiltonian can be 
written as
\bea
   \xi_\lambda(x)&=&(\Omega\delta_\lambda)(x)\nn\cr
  &=&\int_0^1 dy\Bigg(\delta(x-y)
+  {\lambda_B^2\over 8\pi}{\cal P}\Big[{m^2\over y(1-y)}-
{m^2\over x(1-x)}\Big]^{-1} 
{ \tilde A(y)\over \sqrt{x(1-x)y(1-y)}}\Bigg) \delta(y-\lambda)
,\eea
with $\tilde A(\lambda)=\sqrt{\lambda(1-\lambda)}A(\lambda)$.
The left side of the expression gives us the distributional kernel of the
wave operator. (If we are interested in 
incoing pair  we could restrict ourselves to
$\lambda_+$ values).
We can expand an arbitrary vector into
a series of the form $f=\int_0^1 d\lambda f(\lambda)\delta_\lambda(x)$,
and we find
\beq
   \xi(x)=[\Omega(\int_0^1 f(\lambda)\delta_\lambda)](x)=
\int_0^1 d\lambda\Omega(x,\lambda)f(\lambda), \label{trans}
\eeq
and this makes sense in general.
Note that this result is exact (within the linearized large-$N_c$ limit) and 
we have a complete charaterization of the 
set of scattering states once $A(\lambda)$ is given.

We will study the scattering states of the 
coupled equations beyond the bound state tresholds, 
that 
is when the energies are larger than both 
$2m_F$ and $2m_{BR}$. The discussion preceeding these indicates that 
the free part and the interating coupled equations
may not have the same scattering states (see the distributional solution
we present). 
We follow the same idea as in the above  problem
and restrict ourselves to the heuristic Lippmann-Schwinger 
type approach. One can also find the
resolvent exactly and verify the formulae below by more 
careful analysis.

We  solve for the 
scattering amplitude by using a renormalized Lipmann-Schwinger 
equation:
thus we have for the scattering,   
\bea
    \xi_M(x;\Lambda)&=&\delta(x-\lambda_M)+
{\cal P}{1\over \Lambda^2x(1-x)-m_F^2}(1-2x)aA_*(\Lambda)\nn\cr
\xi_N(x;\Lambda)&=&\delta(x-\lambda_N)+{\cal P}{1 \over 
\Lambda_N^2x(1-x)-m_{BR}^2}
\sqrt{x(1-x)}(aB_*(\Lambda)+bA_*(\Lambda))
,\eea
where $A_*(\Lambda), B_*(\Lambda)$ satisfy,
\bea
A_*(\Lambda)&=&{1\over \sqrt{\lambda_N(1-\lambda_N)}}+
(aB_*(\Lambda)+A_*(\Lambda))F(\Lambda,m_{BR})\nn\cr
B_*(\Lambda)&=&{1\over \lambda_M}-{1\over 1-\lambda_M} +
aA_*(\Lambda)\Bigg({\Lambda^2\over m_F^2}-4\Bigg)F(\Lambda,m_F)
,\eea
and $F(\Lambda, m)$ is the same function as  before and 
we choose Max$(4m_{BR}^2, 4m_F^2)<\Lambda^2$, with $\Lambda^2={m_F^2\over 
\lambda_M(1-\lambda_M)}={m_{BR}^2\over \lambda_N(1-\lambda_N)}$.
{\it We can use  $\lambda_M$ as the only  parameter and call it
simply} $\lambda$. 
If we solve for the scattering amplitudes,
\bea
 A_*(\lambda)&=&\Bigg[ 1-\Big({\lambda_R^2\over 4\pi m_F^2}+
{\mu_{2Y}^2\over 4\pi^2m_F^2}\sqrt{(1-4\lambda(1-\lambda))}\ln\Big|
{1-\sqrt{1-4\lambda(1-\lambda)}\over 1+\sqrt{1-4\lambda(1-\lambda)}}\Big|\Big)\nn\cr
&\ &\ \quad \quad \ \ \ 
\times {\lambda(1-\lambda)\over \sqrt{1-4(m_{BR}^2/m_F^2)\lambda(1-\lambda)}}
\ln\Big|{1-\sqrt{1-4(m_{BR}^2/m_F^2)\lambda(1-\lambda)}\over 
1+\sqrt{1-4(m_{BR}^2/m_F^2)\lambda(1-\lambda)}}\Big|\Bigg]^{-1}\nn\cr
&\times&\!\!\!\! \Bigg[{m_F\over m_{BR}}{1\over \sqrt{\lambda(1-\lambda)}}-
{\mu_{2Y}\over 4\pi m_F}{1-2\lambda\over 
\sqrt{1-4(m_{BR}^2/m_F^2)\lambda(1-\lambda)}}
\ln\Big|{1-\sqrt{1-4(m_{BR}^2/m_F^2)\lambda(1-\lambda)}\over 
1+\sqrt{1-4(m_{BR}^2/m_F^2)\lambda(1-\lambda)}}\Big|\Bigg]
.\eea
Note that this result is written in terms of dimensionless
variables.
We can read off $B_*(\lambda)$ as well,
\beq
B_*(\lambda)={1-2\lambda\over \lambda(1-\lambda)}
-{\mu_{2Y}\over 2\pi m_F}\sqrt{1-4\lambda(1-\lambda)}\ln\Big|
{1-\sqrt{1-4\lambda(1-\lambda)}\over 1+\sqrt{1-4\lambda(1-\lambda)}}\Big|
A_*(\lambda)
.\eeq
The reader can check that the
above results actually reduce to the 
phi-four theory results we have found if we set $\mu_{2Y}=0$.

For the sake of completeness we will also present the
scattering solutions for $c_Q(x)$ variables.
Below, we use the same shorthand symbols $A(\lambda)$
and $B_*(\lambda)$ as in the bound state equation 
for $c_Q$:
The renormalized scattering equations become,
\beq
    c_Q(x;\lambda)=\delta(x-\lambda)-{\cal P}{1\over \Lambda^2 x^2-(\Lambda^2-
m_{BR}^2+m_F^2)x+m_F^2}\sqrt{1-x}[(aA(\lambda)+bB_*(\lambda))x+bA(\lambda)]
,\eeq
we should set 
\beq
\Lambda^2={m_F^2\over \lambda}+{m_{BR}^2\over 1-\lambda}
,\eeq
not surprisingly $m_F+m_{BR}\leq \Lambda<\infty$, 
and $A(\lambda), B_*(\lambda)$ found from their definitions,
when we simplify the result it becomes,
\bea
A(\lambda)&=&\Bigg[1-{4\kappa\over \Lambda^2}F_2+
\Big( F_2 {\mu_{1Y}^2\over 64 \Lambda^2}-{\mu_{1Y}^2\over 64\Lambda^2}
(\Lambda^2-m_{RB}^2+m_F^2)F_1+{\mu_{1Y}m_F\kappa\over \Lambda^2}F_1\Big)
\Big( 1+{\mu_{1Y}m_F\over 4}F_1\Big)^{-1}F_2\Bigg]^{-1}\nn\cr
&\times& {1\over \sqrt{1-\lambda}}\Bigg[ 1+{1\over \lambda}
{\mu_{1Y}^2m_F^2\over 32 \Lambda^2
}\Big( 1+{\mu_{1Y}m_F\over 4}F_1\Big)^{-1}F_2\Bigg]\nn\cr
B_*(\lambda)&=& \Big( 1+{\mu_{1Y}m_F\over 4}F_1\Big)^{-1}
\Bigg[{1\over \lambda\sqrt{1-\lambda}}+\Big(8\kappa F_1-{\mu_{1Y}\over 8m_F}
(\Lambda^2-m_{RB}^2+m_F^2)F_1+{\mu_{1Y}\over 8m_F}F_2\Big)A(\lambda)\Bigg].
\eea
where
\beq
   F_2=\ln\Big|{m_{BR}^2\over m_F^2}\Big|,\quad
F_1={1\over u}\ln\Big|{\Lambda^2+m_{BR}^2-m_F^2+u\over 
\Lambda^2+m_{BR}^2-m_F^2-u}\Big|,
\eeq
here $u=\sqrt{(\Lambda^2-(m_F-m_{BR})^2)(\Lambda^2-(m_F+m_{BR})^2)}$.

These define the wave operators of our model, as discussed in the 
simpler model of phi-four coupling. We see that the results are 
fairly complex expressions. One should study various approximate 
forms of these equations, and  a numerical investigation 
of the poles of the amplitudes should give information about 
the bound states.

Our present approach has one weakness, our results are nontrivial 
since the fermions have nonzero mass. This makes it sensitive to the 
sign of the coupling--we treated them as positive, but the 
results are valid if we simply assume them to be 
negative. Physically  more interesting case would be to study the 
massless fermions. If we set $m_F=0$ all the interesting information we have 
is lost in the linear approximation, and we should go beyond this.
This observation suggests that it is necessary to study some 
kind of variational approach to understand the system better.
We plan to investigate this in the future.

\section{Gauged model}

We will assume in this part that the 
model is  gauged by introducing 
$SU(N_c)$ Lie algebra valued gauge potentials $A_\mu$ and refer 
for our conventions to 
\cite{tolyateo}. In 
the light-cone coordinates, and setting $A_-=0$,  the action becomes,
\begin{eqnarray}
 S_{Y}=&&\int dx^+dx^-\bigl[-{1\over 2}\Tr F_{+-}F^{+-}+i\sqrt 
2\psi_L^{*\alpha}\pdr_-\psi_{L\alpha}+
i\sqrt 2\psi^{*\alpha}_R(\pdr_++igA_+)^\beta_\alpha\psi_{R\beta}\nonumber \\
&&-2\phi^{*\alpha}\pdr_-\pdr_+\phi_{\alpha}+ig(\pdr_-\phi^{*\alpha}
{A_+}^\beta_\alpha\phi_\beta-\phi^{*\alpha}{A_+}^\beta_\alpha\pdr_-
\phi_\beta)- m_{B0}^2\phi^{*\alpha}\phi_\alpha 
-{\lambda_{B0}^2 \over 4}(\phi^{*\alpha})^2 \nonumber \\
&&-(\psi^{*\alpha}_L\psi_{R\beta}+\psi^{*\alpha}_R\psi_{L\beta})
(\mu_{1Y}\phi^{*\beta}\phi_\alpha+\mu_{2Y}\phi^{*\lambda}\phi_\lambda
\delta^\alpha_\beta+m_F\delta_\alpha^\beta) 
\bigr] \, .
\end{eqnarray}
The restriction to the color invariant states in the 
gauge theory is actually necessary to make the Hamiltonian 
finite. For this theory the large-$N_c$ limit 
should be a better approximation.
Furthermore one expects baryons in this theory,  
the geometry  of the large-$N_c$ phase space should be 
useful to find a variational ansatz (see \cite{istlect} for 
a nice discussion of these ideas).
Following the same reduction process,
the Hamiltonian becomes    
\beq
  H=H_0+H_Y+H_G
,\eeq
where,
\beq
    H_0={1\over 4}\Big( m_F^2-{g^2\over \pi}\Big)\int {[dp]\over p}M(p,p)+
{1\over 4}\Big(m_{BR}^2-{g^2\over \pi}\Big)\int {[dp]\over |p|}N(p,p),
,\eeq 
$H_Y$ is as given in equation (\ref{hamm}), and the 
gauge contribution is exactly given in \cite{tolyateo},
\bea 
H_G&=&-{g^2\over 16}\int [dpdqdsdt]\ \Big({1 \over (p-t)^2}+{1 \over
(q-s)^2}\Big)\delta[p+s-t-q] M(p,q)M(s,t)\nn\cr
&+&{g^2\over 64}\int [dpdqdsdt]\ 
\Big({1 \over (p-t)^2}+{1 \over (q-s)^2}\Big) 
\delta[p+s-t-q] {qt+ps+st+pq \over \sqrt{|pqst|}} 
N(p,q)N(s,t)\nn\cr 
&+&{g^2\over 8}\int [dpdqdsdt]\ {q+s\over (q-s)^2}{\delta[p+s-t-q]\over
\sqrt{|qs|}} Q(p,q)\bar Q(s,t) 
,\eea

Above we rescaled our coupling constants by a factor of $N_c$
as before  and
 $g^2N_c\mapsto g^2$.
Let us use  exactly the same substitutions as before 
 for the basic variables we have,
and simplify the resulting equations into,
\begin{eqnarray}
   \Lambda_M^2\xi_M(x)&=&\Big(m_F^2-{g^2\over \pi}\Big)
\Big({1\over x}+{1\over 1-x}\Big)\xi_M(x)
-{g^2\over \pi}
\int_0^1 {dy\over (y-x)^2}\xi_M(y)\nn\cr
&-&{\mu_{2Y}m_F\over 4\pi}\Big[ {1\over x}-{1\over 1-x}\Big]\int_0^1
{\xi_N(y)\over \sqrt{y(1-y)}}dy
.\end{eqnarray}
and 
\begin{eqnarray}
\Lambda_N^2\xi_N(x)&=&\Big(m_{RB}^2-{g^2\over \pi}\Big)
\Big({1\over x}+{1\over 1-x}\Big)\xi_N(x)
-{g^2\over 4\pi}\int_0^1{dy\over (y-x)^2}{(x+y)(2-x-y)\over 
\sqrt{x(1-x)y(1-y)}}\xi_N(y)\nn\\
&-&{\mu_{2Y}m_F\over 4\pi\sqrt{(1-x)x}}\int_0^1 dy\Big({1\over y}-{1\over 1-y}
\Big)\xi_M(y)
+{\lambda_R^2 \over 8\pi}\int_0^1dy {\xi_N(y)\over \sqrt{x(1-x)y(1-y)}}
.\end{eqnarray}
Again  we see that the equations for $\xi_M$ and $\xi_N$ are  coupled and
they should be solved together.
\begin{eqnarray}
    \Lambda_Q^2c_Q(x)&=&\Bigg[\Big(m_F^2-{g^2\over \pi}\Big) {1\over x}
+\Big(m_{BR}^2-{g^2\over \pi}\Big){1\over 1-x}\Bigg]c_Q(x)
-{g^2\over 2\pi}\int_0^1{dy\over (y-x)^2}{2-x-y\over \sqrt{(1-x)(1-y)}}
c_Q(y)\nn\\
&-& 8\kappa\int_0^1 {dy\over \sqrt{(1-y)(1-x)}}c_Q(y)
+ {1\over 4} \mu_{1Y}m_F\int_0^1 {dy\over \sqrt{(1-y)(1-x)}}\Big( {1\over x}
+{1\over y}\Big) c_Q(y)
.\end{eqnarray}
These singular integral equations 
can perhaps be investigated numerically.
The linear approximation could be a better one for the
gauged model, since the effect of the gauge interaction is to bring a 
singular operator.
The non-gauged models require renormalizations, it is possible 
that  the above equations will behave better due to the 
singular operators in them.
Another important application is to find a 
variational ansatz for the baryonic solutions and 
have a linear expansion around these solutions.
We plan to study these issues  in more depth in the future.

\section{Acknowledgments}

First of all the author would like to thank Rajeev for 
various discussions about large-$N$ limit field theories.
We also would like to thank M. Arik, J. Gracia-Bondia,
A. Konechny, E. Langmann, J. Mickelsson, T. Rador, and  C. Saclioglu for 
discussions.
The author's stay in Stockholm is made possible by the
G. Gustafsson fellowship, and we are grateful to J. Mickelsson 
for this great experience.
Our work is also partially supported by the Young Investigator  Award
of Turkish Academy of Sciences (TUBA-GEBIP).

\section{Appendix: Reduction of the Hamiltonian}

In this appendix we will give some of the details of 
the reduction of the Hamiltonian to the desired color invariant
products. 
Let us recall that the  Hamiltonian is given in 
equation number (\ref{hamm}). 
Let us start with the term,
\beq
   {\sqrt{2}\over 2}\mu_{1Y}^2\int dx^-\ \hat \psi^{\dag\beta}\hat\phi_\beta
\hat\phi^{\dag\alpha}{1\over i\partial_-}\hat \phi_\alpha\hat\phi^{\dag\lambda}
\hat \psi_\lambda
.\eeq
When we write this in terms of the Fourier mode expansions, 
it  becomes,
\bea
&\ &-{\mu_{1Y}^2\over 8}\int {[dkdldtdpdqds]\over \sqrt{|stkq|}} 
{\delta[k-l+t-s+q-p] \over t-k+l}
\chi^{\dag\alpha}(p)a_\alpha(q):a^{\dag\beta}(s)a_\beta(t):a^{\dag\lambda}(k)
\chi_\lambda(l)\nn\cr
&\ &- {\mu_{1Y}^2N_c\over 16}\int {[dkdldqds]\over \sqrt{|qk|}}
\chi^{\dag\alpha}(p)a_\alpha(q)a^{\dag\lambda}(k)\chi_\lambda(l)\delta[k-l+q-s]
\Big(\int [ds]{\sgn(s)-1\over s(k-l-s)}\Big)
.\eea
Notice that the divergent integral is isolated and a 
principal value regularization calculation shows that
\beq
{\cal P}\int [ds] {\sgn(s)-1\over s(k-l-s)}={1\over \pi(k-l)}\ln\Big|{k-l\over 
\epsilon_R}\Big|
,\eeq
where 
$\epsilon_R$ is an infrared cut-off.
If we are only allowed to introduce  local
counter terms {\it in the original action} we should introduce a 
momentum scale $\mu_R$ so that we can 
separate the momentum dependent part and purely divergent 
part of these type expressions:
(this point  is somewhat difficult to decide   for this particular term since 
it is not possible to write such a term  in the original action)
\beq
{\cal P}\int [ds] {\sgn(s)-1\over s(k-l-s)}={1\over \pi(k-l)}\ln\Big|{k-l\over 
\epsilon_R}\Big|={1\over \pi(k-l)}\Bigg(\ln\Big|{k-l\over \mu_R}\Big|
+\ln\Big|{\mu_R\over \epsilon_R}\Big|\Bigg)
.\eeq    
If we remove the divergent part by 
a counter term of the 
form 
\beq
\Bigg({\mu_{1Y}^2N_c\over 16\pi}\ln
\Big|{\mu_R\over \epsilon_R}\Big|+8\kappa_R(\mu_R)\Bigg)
\int {[dpdqdkdl]\over \sqrt{qk}}{1\over k-l}\delta[p-q+k-l]
\chi^{\dag\alpha}(p)a_\alpha(q)a^{\dag\beta}(k)\chi_\beta(l)
.\eeq
the finite term comes out  to be
\beq
\int{[dkdldqds]\over \sqrt{|lq|}}\Big[8\kappa_R(\mu_R)-{\mu_{1Y}^2\over 8\pi}
\ln\Big|{k-l\over \mu_R}\Big|\Big]
{1\over q-s}\chi^{\dag\alpha}(k)a_\alpha(l)a^{\dag\beta}(q)\chi_\beta(s)
\delta[k-l+q-s]
.\eeq
If we require  the theory not to have  a dependence on the 
arbitrary scale we introduced, it is 
natural to demand that the residual coupling 
to vary  under a change of scale according to
\beq
\kappa_R(\mu_R)=\kappa_R(\tilde\mu_R)-{\mu_{1Y}^2\over 64\pi}\ln\Big|
{\mu_R\over \tilde\mu_R}\Big|
.\eeq
The reader may be alarmed by the 
nonlocal expression in the interaction, but if we actually go
back to the position space,
the inverse Fourier transform gives a term, upto some constants, 
\beq 
 \int dx dy \Big( a\sgn(x-y)-{\ln|\mu_R(x-y)|\sgn(x-y)}\Big)
\hat \psi^{\dag\alpha}(x)\hat \phi_\alpha(y)\hat \phi^{\dag\beta}(y)
\hat \psi_\beta(x)
,\eeq
which has a logarithmic 
correction to the sign function.
(This behaves worse for the short-distance than the coulomb 
potential $|x-y|$, but we should interpret $\sgn(x-y)= 0$ if $x=y$, so
there is no singularity at the short distance).
In the text we will only consider the cases where these non-local terms
dropped, or removed by taking them as part of the counter terms
in the action.

(Notice that this is the interaction  one finds 
if we use a parity broken model as 
in \cite{tolyateo}. The sign of the remaining 
interaction term is not determined since it
is not in the original action it should be left as an  arbitrary
parameter).

As another example we will discuss the term,
\beq
  {\sqrt{2}\over 2}{m_F\mu_{2Y}}\int dx^-  \ \hat \psi^{\dag\alpha}
{1\over i\partial_-}\hat \psi_\alpha :\hat \phi^\sigma\hat\phi_\sigma:
,\eeq
a Fourier expansion and removing the vacuum expectation value 
gives us a term 
\bea
   &-&{\mu_{2Y}m_F\over 4}\int {[dpdqdsdt]\over \sqrt{|st|}}
{\delta[p-q+s-t]\over s-t-q}:\chi^{\dag\alpha}(p)\chi_\alpha(q):
:a^{\dag\beta}(s)a_\beta(t):\nn\cr
&+&{\mu_{2Y}N_cm_F\over 8}\int {[ds]\over |s|}:a^{\dag\beta}(s)a_\beta(s):
\int [dp] {1-\sgn(p)\over p}
,\eea
and 
\beq
   {\cal P} \int [dp] {1-\sgn(p)\over p}=-2
\ln\Big|{\Lambda_R\over \epsilon_R}\Big|
,\eeq
where we have $\epsilon_R$ and $\Lambda_R$ as the infrared and ultraviolet 
cut-offs respectively.
If we introduce a boson mass counterm of the form
\beq
{\mu_{2Y}N_cm_F\over 4}\ln\Big|{\Lambda_R\over \epsilon_R}\Big|
\int {[ds]\over |s|}:a^{\dag\beta}(s)a_\beta(s):
,\eeq
 this term will be cancelled.

The other terms are also done in the same way, and the rest is to collect 
all terms to find the Hamiltonian in terms of large-$N_c$ bilinears.

\end{document}